\documentstyle[pra,aps,preprint]{revtex}
\begin{document}
\draft
\title{Optical Spectra in the Ferromagnetic States \\
near the Charge Ordering}
\author{S. Ishihara, and N. Nagaosa}
\address{
Department of Applied Physics, University of Tokyo,
Bunkyo-ku, Tokyo 113
}
\date{\today}
\maketitle
\begin{abstract}
The optical conductivity is studied numerically 
for the ferromagnetic 
metallic state close to the charge ordering  
observed in perovskite manganites. 
We focus on roles of the charge fluctuation on the incoherent 
component in the optical spectra, and 
consider the Coulomb interaction in the 
one- and two-dimensional spinless fermion models. 
The numerical results show that 
the incoherent part is not noticeable, 
and near the quarter-filling, 
most of the spectral weights are transferred to the Drude component.  
In the one dimensional model, there is little or no spectral weight 
in the energy region between 
the Drude weight and the remnant of the absorption edge which is 
pushed up with the carrier doping, in contrast to the two dimensional one.  
The discrepancy between our calculations and the experiments 
suggests that the orbital degrees of freedom play important roles. 
\
\end{abstract}
%
\par
\eject
\narrowtext

After the discovery of High-Tc cuprates, 
the electronic structures in many  
transition-metal oxides have been reinvestigated intensively 
from the renewed viewpoints. 
One of the attractive materials 
is manganese oxides with the perovskite structure 
$\rm A_{1-x}B_xMnO_3$ $\rm (A=La, Pr,Nd, \  B=Sr,Ca)$
owing to the negative colossal magnetoresistance (CMR) and its potential 
applications \cite{cha,hel,toku}. 
The parent compound, $\rm LaMnO_3$, is an antiferromagnetic 
insulator \cite{wk,mat}, where  
the valence of the Mn ion is +3  
and the electronic configuration is 
$(t_{2g})^3(e_g)^1$ 
because of the strong Hund coupling. 
It is noted that, 
as well as the spin and charge, 
the $e_g$ electron in $\rm Mn^{3+}$ ion has the 
orbital degrees of freedom. 
The layer-type antiferromagnetic spin alignment and 
the two-dimensional character of the spin wave \cite{hiro} 
are attributed to 
the orbital ordering accompanied with the Jahn-Teller lattice distortion
\cite{good,kan,ishi}.  
As the concentration of the divalent cation is increased, 
the system turns to the ferromagnetic metallic (FM) states  
at $x \sim 0.175$ in $\rm La_{1-x}Sr_xMnO_3$, 
and CMR is observed near the ferromagnetic Curie temperature ($T_c$). 
With further doping, 
$T_c$ is gradually decreased    
and the charge ordered (CO) phase comes out around  
$x \sim 0.5$ in $\rm Nd_{1-x}Sr_xMnO_3$ \cite{kuwa} 
and $\rm La_{1-x}Ca_xMnO_3$ 
\cite{wk}. 
In the CO phase, the spin and orbital orderings 
coexist. 
Therefore, it is recognized that 
the perovskite manganites are one of the typical materials where  
the spin, charge and orbital degrees of freedom are strongly correlated. 
\par
In the FM phase, 
some anomalous features in the optical conductivity $\sigma(\omega)$
are recently reported \cite{oki}. 
$\sigma(\omega)$ shows the two-component feature, that is, 
the sharp Drude peak with the small intensity and the broad 
incoherent band up to about 1eV. 
Even at the lowest temperature, where 
the magnetization is almost saturated, 
the integrated intensity of these components   
is still growing up with decreasing the temperature. 
Because the spins are perfectly aligned at that temperature,
the results indicate that  
another degrees of freedom are still active. 
In our previous paper \cite{ishi2}, 
we focused on the orbital degrees of freedom in perovskite 
manganites and investigated the nature of the orbital fluctuation 
in the FM phase. 
With considering the strong electron-electron interactions 
and the anisotropy of the electron transfer, 
it is concluded that the long range orbital ordering 
is prevented by the two-dimensional character of the orbital fluctuation. 
We proposed the orbital liquid states 
where the orbital fluctuation is still active down to low temperatures.  
The numerical calculation based on this scenario 
shows the broad spectra in the optical conductivity. 
It has been proposed that the orbital 
liquid state is one of the promising candidates 
for the origin of the incoherent spectra observed in perovskite manganites. 
\par
As well as the orbital degrees of freedom, the Mn ion in the FM 
phase has the charge degrees of freedom, as was pointed out above. 
In the CO phase around $x \sim 0.5$, 
the charge excitation gap, showing the gentle slope, is found in the optical 
conductivity in $\rm Pr_{0.6}Ca_{0.4}MnO_3$ \cite{oki2}. 
Introducing the charge carriers or applying the magnetic 
field to the CO state, 
the charge ordering easily melts and it turns into  
the FM phase at low temperatures. 
In the FM phase near the charge ordering, 
the charge fluctuation is expected to be large 
due to the competition between the 
electron kinetic energy and the long range Coulomb interaction,  
and may bring about the incoherent component. 
Therefore, it is important to study to what extent the 
remnant of the charge ordering is 
relevant to the optical spectra. 
\par
In this letter, 
we study numerically the optical conductivity 
in the FM phase near the charge ordering. 
We focus our attention to the roles of 
the charge degrees of freedom,  
because the orbital degrees of freedom have been 
investigated in our previous paper \cite{ishi2}.  
We employ the exact diagonalization 
of the finite size systems, 
and conclude that its 
contribution is less dominant on the incoherent broad band 
in the region $x < 0.4$ of our interest. 
\par
We consider the $e_g$ electron 
with the single orbital in the perfectly spin 
polarized ferromagnetic state. 
In order to describe the fermionic excitations in this phase, 
we introduce the spin- and orbital-less fermion in each Mn sites. 
The on-site Coulomb interactions between the $e_g$ electrons and the 
Hund coupling between the $e_g$ and $t_{2g}$ spins are irrelevant,  
while the inter-site Coulomb interaction dominates 
the charge ordering and its fluctuations in this case. 
We consider the following Hamiltonian,
\begin{equation}
H=t\sum_{<ij>} (c_i^\dagger c_j+H.c)
 +V\sum_{<ij>} n_i n_j  \ , 
\label{eq:ham}
\end{equation}
where $c_i^\dagger$ is the spin- and orbital-less fermion operator 
at site $i$, 
and $n_i$ is the charge density operator 
defined as $n_i=c_i^\dagger c_i$. 
$t$ and $V$ is the transfer integral for the fermion and 
the Coulomb interaction between the nearest neighbor sites, 
respectively. 
In order to obtain the ground state wave function and 
the optical conductivity spectra, 
we utilize the exact diagonalization method 
i.e., the Lanczos method and the recursion method 
in the finite size system \cite{gag,nis}. 
With considering the symmetry of the lattice and the capacity 
of the memory in the computer, 
the one dimensional lattice with $N_L=16 \times 1$ sites and 
the two dimensional one with $N_L=4 \times 4$ sites 
are studied. 
The size effects are studied 
in the one dimensional lattice, 
and the convergence of the results is confirmed in the $16 \times 1$ 
sites case. 
The periodic boundary condition is adopted 
to avoid the fictitious low energy excitation near the free boundaries.  
\par
In general, 
the optical conductivity is decomposed into the two parts, i.e.,  
the Drude component and the regular part, as follows, 
\begin{equation}
\sigma(\omega)=D\delta(\omega)+\sigma^{reg}(\omega) \ , 
\label{eq:sigma} 
\end{equation}
where $\sigma^{reg}(\omega)$ is defined by 
\begin{equation}
\sigma^{reg}(\omega)= {\pi e^2 \over N_L} \sum_{n \ne 0}
{ | \langle n | j | 0 \rangle |^2 \over E_n-E_0} 
\delta(\omega-E_n+E_0) \ . 
\label{eq:reg}
\end{equation}
In Fig. 1, we present the doping dependence of the regular part of the 
optical conductivity spectra in the one and two dimensional cases. 
The Drude part will be discussed later. 
In the case where the ground state is degenerate, 
the spectra are averaged in the momentum space. 
The value of $V/t$ is chosen as 12 and 4 in the one 
and two dimensional case, respectively, 
in order to true up the absorption edges. 
Although these values are taken to be larger in comparison 
with that in the actual compounds,
this is for the purpose to avoid the finite size effects. 
The fermion concentration $n=N/N_L$ is identified with 
$x$ in manganites, and $n=0.5$ corresponds to the half-filling.  
In Fig. 1, the optical gap is clearly shown at $\omega \sim V$ 
and at $\omega \sim 3V$ in the one and two dimensional case, respectively. 
This is nothing but the charge ordered insulator, 
where the charge alternation occurs in the 
one and two dimensional lattices. 
The optical gap corresponds to the charge transfer gap, 
because an transfer of the fermion requires the energy 
of $V$ and $3V$ in the one and two dimensional cases, respectively. 
With decreasing the number of the fermion, 
the system departs from the commensurate charge ordered phase. 
From the results of the calculated Drude weight, 
it is concluded that the system is metallic, except for the half-filling case. 
As the system departs from $n=0.5$, 
in the one dimensional case (Fig. 1(a)), 
the energy at the absorption edge is gradually increased 
with the reduction of its intensity. 
There is little or no spectral weight between the Drude part and the 
remnant of the absorption edge above $V$, 
and the spectral weight is transferred from the later to the former 
with doping.  
This picture for the optical spectra in the one dimensional chain 
are consistent with results predicted by the bosonization 
procedure \cite{mori}, where the above two components in the spectra 
is interpreted as the intra- and inter-band transitions at 
the Luther-Emery point. 
Near the quarter-filling ($n=0.25$) in Fig. 1(a), 
the Drude part governs the almost all spectral weight 
and there are no remarkable weight in the finite energy region. 
On the other side, 
in the two dimensional $4 \times 4$ lattice, 
the charge gap is collapsed and the spectral weight appears 
within the gap with doping, in contrast 
to the one dimensional case where the remnant of the
absorption edge is pushed up.  
This is because the charge excitations at $\omega \sim V$ and 
$\omega \sim 2V$ become possible by doping. 
However, we cannot find any remarkable spectra in 
the region of $\omega < V$ in Fig. 1(b). 
At the quarter-filling, only the small amount of the spectra 
remains in the finite energy region similarly to 
the one dimensional case. 
\par
Next we study the doping dependence of the Drude weight. 
In Fig. 2, we show  
the Drude weight ($D$) and the total spectral weight ($f$), 
in one- and two-dimensional cases. 
The Drude weights in the finite size cluster with the 
periodic boundary condition 
are numerically estimated by the following relation \cite{sca,dag}: 
\begin{equation}
{D \over e^2}=-{\pi \over N_L} <T> 
  -{2 \over e^2 N_L} \sum_{n \ne 0}   
{ | \langle n | j | 0 \rangle |^2 \over E_n-E_0} \ , 
\label{eq:drud}
\end{equation}
where 
the first term in the r.h.s is the total spectral weight $f$, and 
$<T>$ is the expectation value of the kinetic energy in the 
ground state. 
The validity of the above formula has been carefully 
examined in the one dimensional 
Hubbard model by the several authors \cite{sca,dag}. 
It has been confirmed that 
for $N \rightarrow \infty$, $D$ estimated by Eq. (\ref{eq:drud}) is 
close to the Bethe-ansatz value, and its doping dependence 
is qualitative correct even in the small cluster system. 
At half-filling in Figs. 2(a) and (b), 
we find small positive and 
negative values of $D$  
being attributed to the finite size effects. 
As the several authors have mentioned \cite{sca,dag},  
it becomes inconspicuous with increasing the size and 
the magnitudes of the interaction. 
With doping into the insulating states at the half-filling,   
the Drude weight appears and the system becomes metallic.  
In the one dimensional case, $D$ is rapidly growing up 
with the carrier density $|0.5-n|$, 
and its value approaches to $f$, immediately. 
With further doping, 
both $D$ and $f$ go toward zero at $n=0.0$ and 1.0 cases. 
On the other hand, in the two dimensional lattice, 
$D$ is suppressed near the half-filling,  
reflecting the spectral weight 
in the finite frequency region, as shown in Fig. 1(b). 
As the system approaches to the quarter-filling case, 
however, almost all the weights go to the 
Drude component.  
With increasing $V/t$, the rising of $D$ and $f$ around the half-filling 
become slower, and the Drude part dominates the total spectra 
more and more. 
\par
In Fig. 3, we present the static 
charge correlation function $\chi_c(\vec k)$ 
calculated in the finite size system. 
$\chi_c(\vec k)$ is  
defined by $\chi_c(\vec k)=F.T.<\delta n_i \delta n_j>$
with $\delta n_i=n_i-n$. 
At half-filling,  
$\chi_c(\vec k)$ has a sharp peak at $\pi/a$ and $(\pi/a, \pi/a)$  
in the one and two dimensional lattice, respectively,
reflecting the commensurate charge ordering. 
With doping carriers, 
the intensity of $\chi_c(\pi/a)$ in the one dimensional lattice 
is drastically decreased 
and the broad incommensurate peak appears around $k=\pi/a-\delta$. 
In the cases of $ n \ge 0.4375$ (more than one hole doping), 
the magnitude of $\chi_c(k)$ is smoothly reduced with the doping.   
On the other hand, in the two dimensional $4 \times 4$ lattice, 
the noticeable peak at $(\pi/a,\pi/a)$ still remains up to 
the case of $n=0.375$. 
With further doping, the intensity of the 
correlation function is almost saturated, 
and the incommensurate peak appears at $(\pi/2a,\pi/a)$ at 
the quarter filled case. 
The doping dependencies of $\chi_c(\pi/a)$ and $\chi_c(\pi/a,\pi/a)$ 
are strongly correlated with these of the Drude weight, 
shown in Fig. 2, both in the one and two dimensional cases, i.e., 
as the charge correlation at the boundary in the Brilloun zone 
is diminished, the Drude weight is gradually growing up.  
It is suggested that the strong charge correlation 
prevents the coherent motion of the doped carrier.  
\par
As we show in Figs. 1 and 2, 
any conspicuous spectral weight are not found 
in the region of $\omega < V$ 
in both one and two dimensional cases. 
This feature is quite in contrast to the results 
for the positive-U Hubbard and t-J models \cite{ste,mor,fye} which 
has been extensively studied 
in connection with the mid-infrared absorption observed 
in High-Tc cuprates.
As the carriers are doped in the half-filled 
antiferromagnetic insulator, 
the spectral weight appears around $\omega \sim J \sim t^2/U$, 
in addition to the $\omega \sim U$ structure originated from 
the transition between the Hubbard bands in the Hubbard model. 
In the two dimensional case, 
this structure becomes remarkable and its weight is comparable 
to that located around $U$. 
The origin of this mid-infrared component 
is interpreted as the incoherent motion of the 
doped holes which creates 
distortions in the antiferromagnetic background. 
In a word, the presence of the spin degrees of freedom 
is essential to the emergence of the band. 
On the contrary, 
only the charge degrees of freedom is 
taken into account in the present model, 
describing the perfectly polarized ferromagnetic states 
without the orbital degeneracy. 
There are not additional degrees of freedom which 
govern the optical spectra smaller than $V$. 
\par
Turning to the experimental results,  
the frequency dependence of the conductivity spectra 
is measured in $\rm La_{1-x}Sr_xMnO_3$ 
at x=0.0 (antiferromagntic insulator) \cite{ari}, 
0.1(ferromagnetic insulator), 0.175(FM) and 0.3(FM) \cite{oki}. 
The weight and width of the incoherent band 
becomes the most intense at x=0.175 
and it entirely fills up the optical gap (about 1eV) 
observed at x=0.0 and 0.1. 
With further hole doping (x=0.3), the weight is rather diminished 
in spite that the charge fluctuation is expected to be large, 
as the system approaches to x=0.5. 
The incoherent band also appears by applying the magnetic field in 
$\rm Pr_{0.6}Ca_{0.4}MnO_3$ where the charge ordering is realized 
without the magnetic field. \cite{oki2}. 
At about 6.5T, 
the spectra are continuously spread up to about 0.8eV and 
entirely covers up the charge gap observed at zero magnetic field. 
On the contrary, in the present theoretical calculation, 
the almost all weights are transferred to the Drude part
near $n=0.25$ where the charge correlation at the zone boundary 
are fairly weaken, and 
even around the half-filling, 
there is no noticeable spectra in $\omega < V$. 
Although the numerical calculations are limited 
in the one and two dimensional 
finite size systems, 
we have no doubt that 
the contributions from the charge fluctuation 
are less dominant on the incoherent component 
observed in the perovskite manganites. 
\par
The results presented in this paper suggest the other candidates 
for the origin of the broad incoherent component 
in the optical spectra, such as the low dimensional orbital fluctuation 
proposed by the present authors \cite{ishi2}, 
the interband transition proposed by Shiba. et.al. \cite{shiba} 
and so on. 
Actually, incoherent band based on the orbital fluctuation 
are calculated by the static approximation \cite{ishi2} where 
the orbital fluctuation is treated as the slowly varying degrees 
of freedom. 
It is shown that 
the width of the spectra is an order of the transfer integral  
and its weight is expected to be growing up with decreasing the 
temperature. 
Also, our previous calculation is consistent with the 
observed two-component 
feature in the optical conductivity with considering the 
interaction between the orbital and Jahn-Teller distortion. 
\par
\medskip
\noindent
ACKNOWLEDGMENT
\par
\noindent
The authors would like to thank Profs. Y.Tokura, S.Maekawa, H.Shiba, and 
E.Dagotto 
for their valuable discussions. 
One of the authors (S.I.) would like to express his thanks to 
Drs. W.Koshibae, and K.Tsutsui for their comments on   
the numerical calculation. 
This work was supported by Priority Areas Grants from the Ministry of 
Education, Science and Culture of Japan,  
the Proposal-Based Advanced Industrial Technology 
R$\&$D Program from the New Energy and 
Industrial Technology Development Organization (NEDO) of Japan, 
and Grant-in-Aid for COE research.  
The part of the numerical calculation is performed in 
the HITACS-3800/380 supercomputing facilities in Institute for 
Materials Research, Tohoku University. 
\vfill
\eject
\noindent
Figure captions
\par 
\medskip
\noindent
Fig. 1: 
 The optical conductivity spectra in the one dimensional 16 sites 
(Fig. (a)) and the two dimensional $4 \times 4$ sites (Fig. (b)).  
$V/t$ is chosen as $V/t=12$ and $V/t=4$ in the one and two dimensional case, 
respectively. 
The calculations are performed at $\omega+i \delta$ with 
$\delta/t=0.0002$. 
\par
\par
\medskip
\noindent
Fig. 2: 
The Drude weight $(D /e^2)$ and the total spectral weights $(\pi<T>/N)$ 
in the one dimensional 16 sites (Fig. 2(a)) and the two dimensional 
$4 \times 4$ sites (Fig. 2(b)).
\par
\par
\medskip
\noindent
Fig. 3:  
The charge correlation function 
$\chi_c(\vec k)=F.T.<\delta n_i \delta n_j>$ 
in the one dimensional 16 sites (Fig. 3(a)) and the two dimensional 
$4 \times 4$ sites (Fig. 3(b)). 
The parameter is chosen as $V/t=4$ in the both cases. 
\par
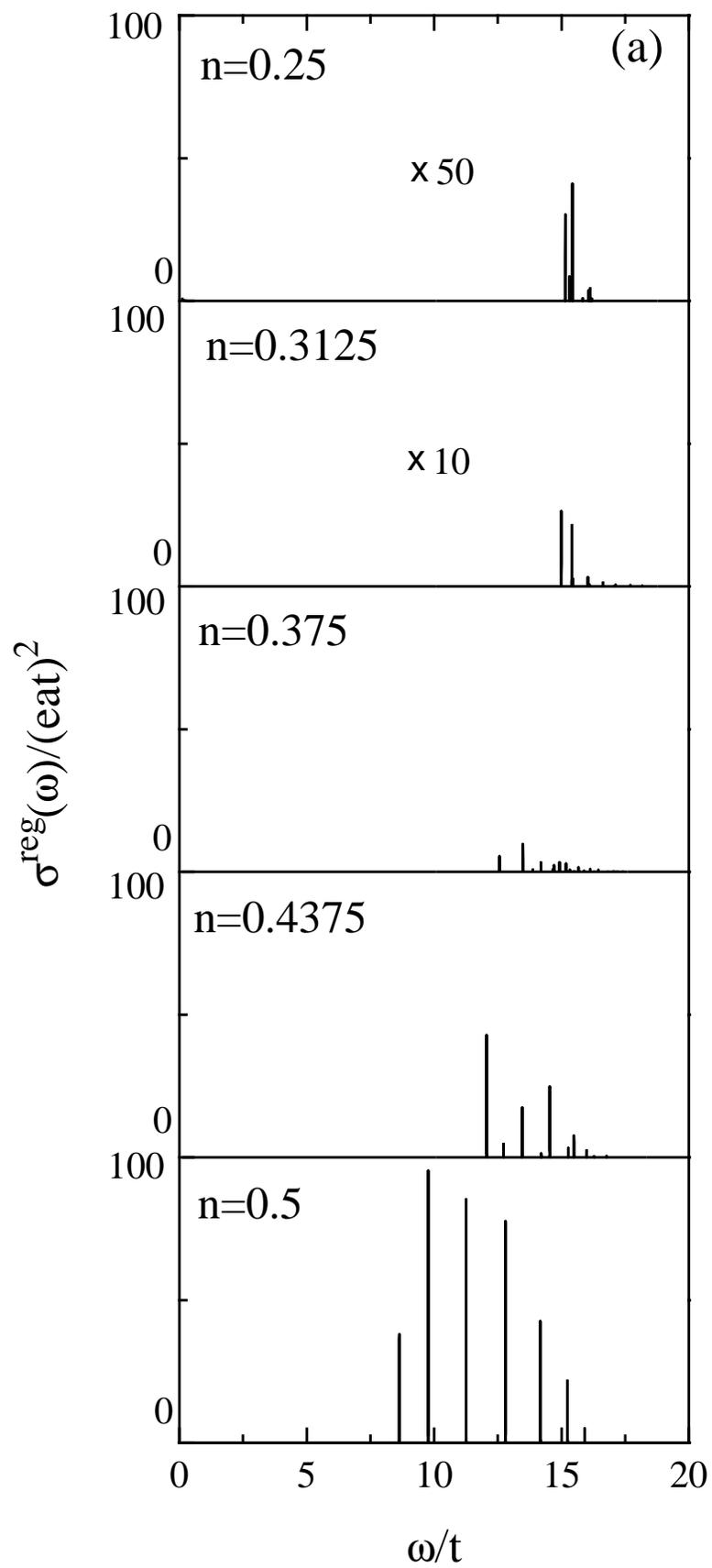
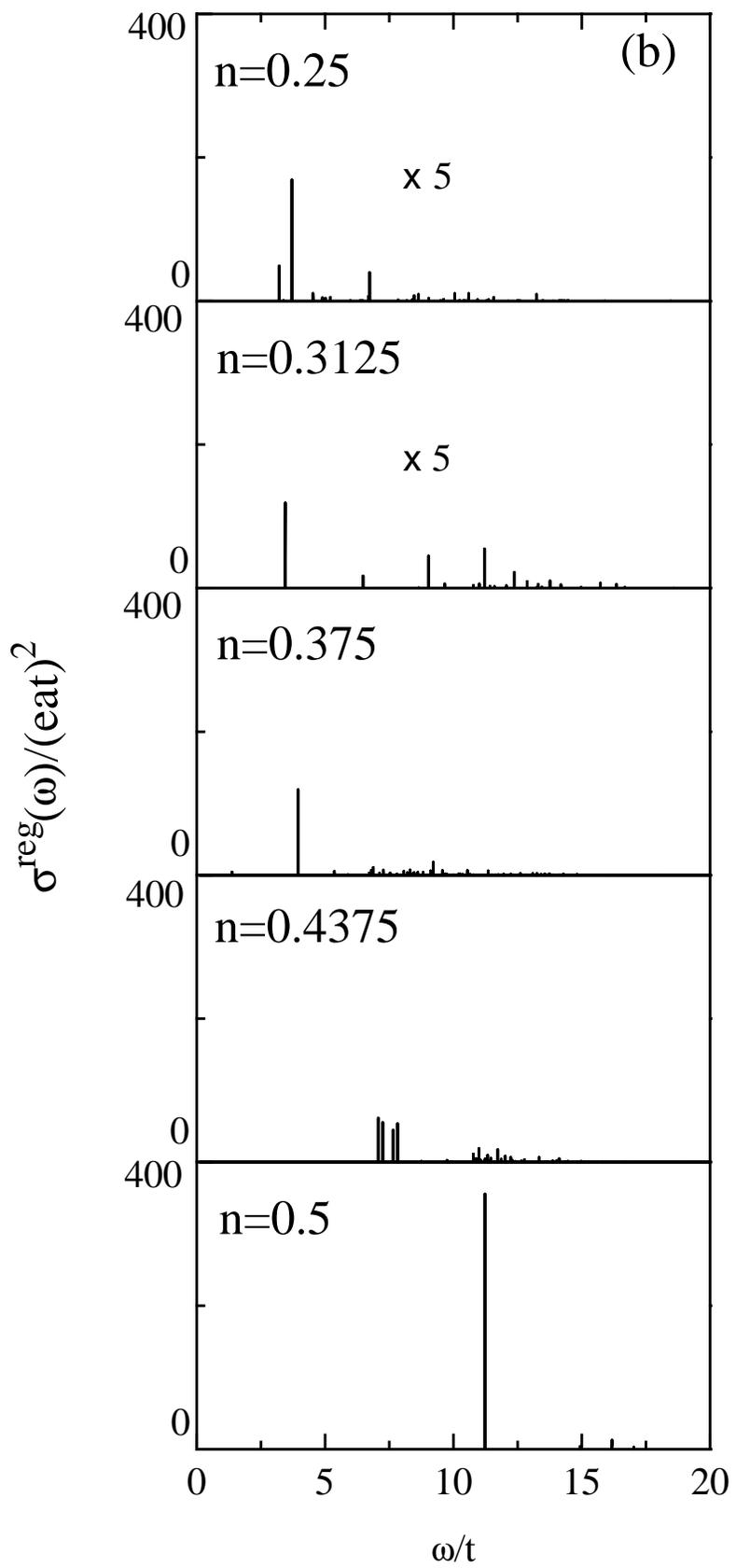
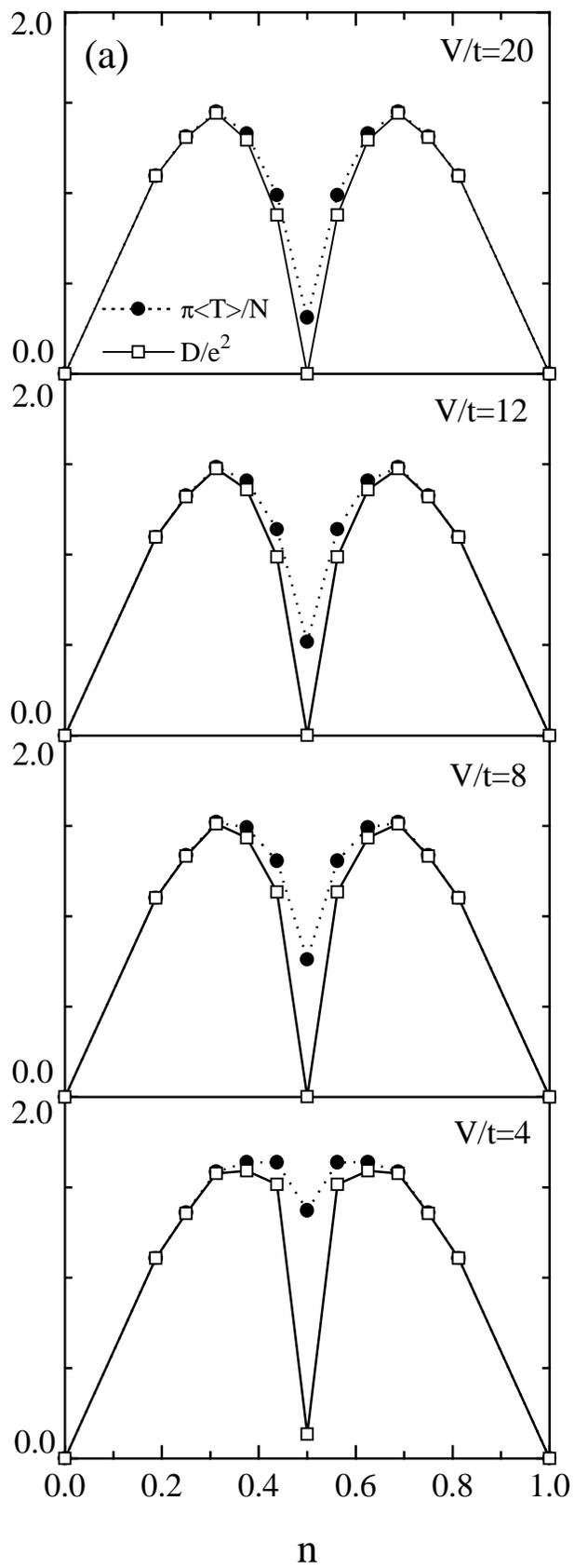
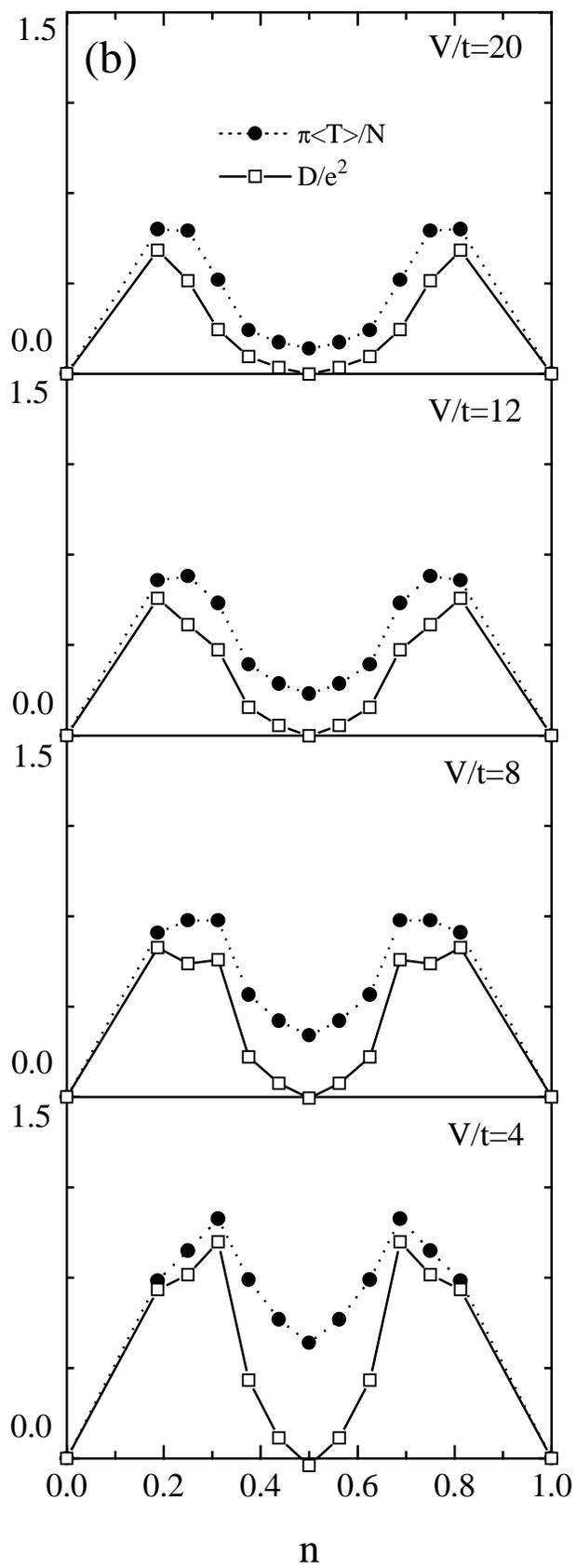
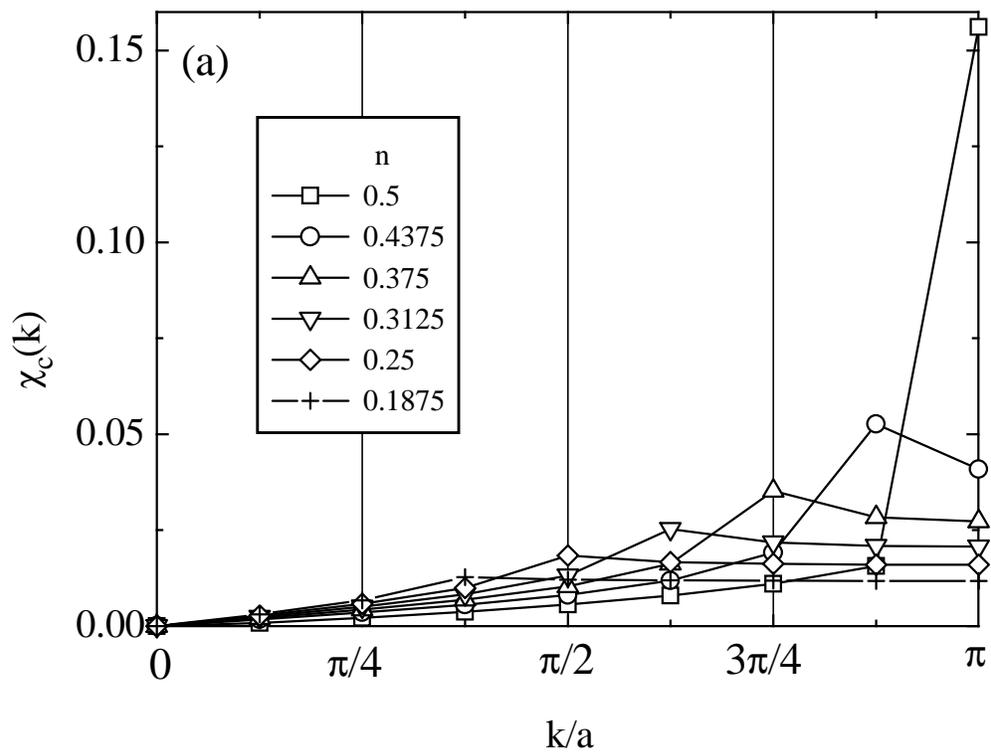
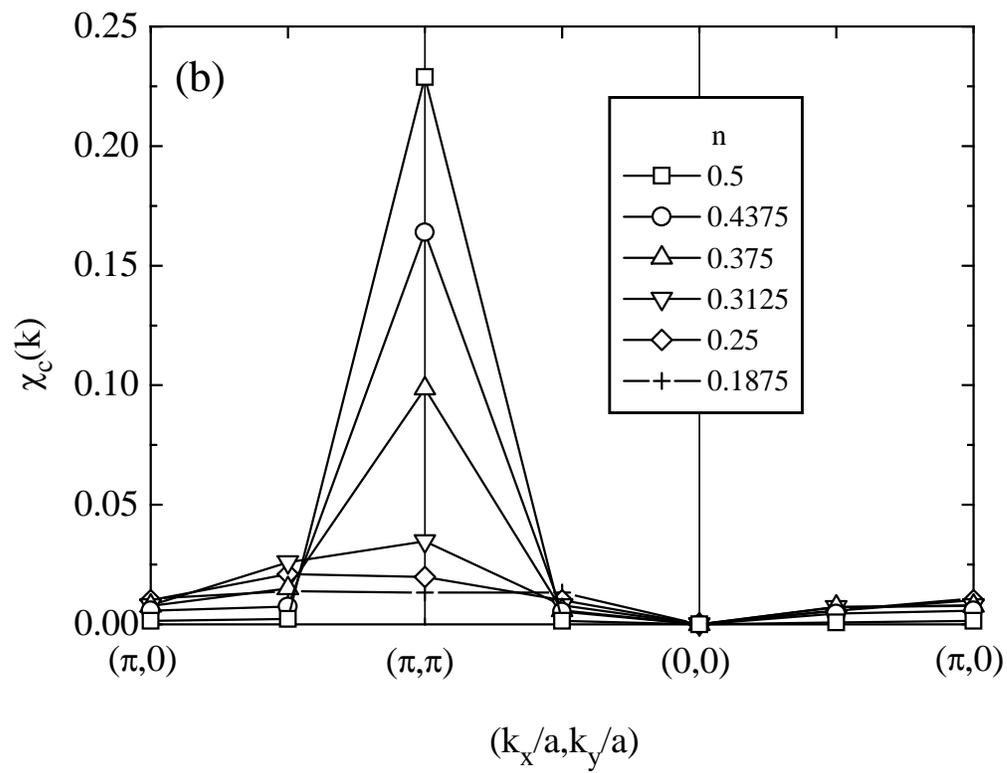
\end{document}